\documentclass[letter]{aa}
\usepackage{txfonts}
\usepackage{epsfig,graphics,longtable,amssymb,verbatim}
\usepackage{natbib}
\bibpunct{(}{)}{;}{a}{}{,} 

\begin{document}

\title{On the H$\alpha$ emission from the \object{$\beta$ Cephei} system}

\author{R.~S.\ Schnerr\inst{1}
      \and H.~F.\ Henrichs\inst{1}
      \and R.~D.\ Oudmaijer\inst{2}
      \and J.~H.\ Telting\inst{3}
}

\institute{Astronomical Institute ''Anton Pannekoek``, University of Amsterdam, Kruislaan 403, 1098 SJ Amsterdam, Netherlands
  \and School of Physics and Astronomy, EC Stoner Building, University of Leeds, Leeds LS2 9JT, UK
  \and Nordic Optical Telescope, Apartado 474, 38700 Santa Cruz de La Palma, Spain
}

\offprints{R.S. Schnerr,
\email{rschnerr@science.uva.nl}}

\date{Received date / Accepted date}


\abstract{Be stars, which are characterised by intermittent emission in their hydrogen lines, are known to be fast rotators. This fast rotation is a requirement for the formation of a Keplerian disk, which in turn gives rise to the emission. However, the pulsating, magnetic B1IV star \object{$\beta$ Cephei} is a very slow rotator that still shows H$\alpha$ emission episodes like in other Be stars, contradicting current theories.}
{We investigate the hypothesis that the H$\alpha$ emission stems from the spectroscopically unresolved companion of $\beta$ Cep.}
{Spectra of the two unresolved components have been separated in the 6350-6850\AA\ range with spectro-astrometric techniques, using
11 longslit spectra obtained with ALFOSC at the Nordic Optical Telescope, La Palma.}
{We find that the H$\alpha$ emission is not related to the primary in $\beta$ Cep, but is due to its 3.4 magnitudes fainter companion. This companion has been resolved by speckle techniques, but it remains unresolved by traditional spectroscopy. The emission extends from about $-$400 to +400 km~s$^{-1}$. The companion star in its 90-year orbit is likely to be a classical Be star with a spectral type around B6-8.}
{By identifying its Be-star companion as the origin of the H$\alpha$ emission behaviour, the enigma behind the Be status of the slow rotator $\beta$ Cep has been resolved.}

\authorrunning{R.S.\ Schnerr et al.}
\maketitle

\section{Introduction}
The well-known pulsating star \object{$\beta$ Cephei} (\object{HD 205021}) has been classified as B1IVe. Its Be status was assigned after the star showed prominent emission in H$\alpha$. The presence of this emission has been reported from time to time since 1933 \citep{karpov:1933}, but often the emission disappeared or was not noticed. A new H$\alpha$ emission episode was discovered in 1990 \citep{mathias:1991,kaper:1995}, which decayed in about 10 years. \citet{neiner:2001} found that the emission was back again within several years. A summary of the emission phases until 1995 is given by \citet{panko:1997}.

This behaviour is typical of Be stars. The enigma is that nearly
all Be stars are rapid rotators with equatorial rotation rates of
typically $\sim$70-80\% of the critical rotation velocity
\citep[e.g.][]{porter:2003}, or perhaps even higher
\citep{townsend:2004}. However, $\beta$ Cep is a very slow rotator
with $v \sin i \approx$ 25 km~s$^{-1}$ and has a very well-determined rotation period of 12.00 days \citep{henrichs:1993}, much
longer than the inferred rotation periods of other Be stars.
Interestingly, the star was discovered to be an oblique magnetic
rotator \citep{henrichs:2000a} with a polar field of $\sim$360 G
\citep[see also][]{donati:2001}, which strongly modulates the outflowing stellar wind with the rotation period. This has been very clearly observed in the UV resonance lines of \ion{C}{iv}, \ion{Si}{iv}, and \ion{N}{v} with the $IUE$ satellite over more than 15 years. This spectral line modulation could be modelled reasonably well as being due to the interaction of the magnetic field with the stellar wind \citep{schnerr:2006d}, similar to the rotationally modulated winds of the magnetic Bp stars
\citep[e.g.][]{townsend:2005}, which also show H$\alpha$ emission.

The serious problem, however, is that every model so far predicts
that this 12-day rotation period of $\beta$ Cep should also be
clearly visible in the H$\alpha$ emission (probing the outflow near
the stellar surface), whereas no sign of any 12-day modulation could be found
in more than 300 high-resolution H$\alpha$ profiles taken over 6
years \citep{henrichs:2006}. This discrepancy seriously hampers our
understanding of the Be phenomenon: if $\beta$ Cep really belongs to the (phenomenologically defined) class of Be stars, rapid rotation would not be required for the explanation of the Be phenomenon, opposed to all existing models. In addition, the origin of the unmodulated H$\alpha$ emission would remain a mystery. Current modelling efforts would clearly benefit from resolving this critical issue.

The aim of this study is to investigate the hypothesis that the source of the H$\alpha$ emission is not $\beta$ Cep itself, but its nearby companion, which has been resolved by speckle techniques. This suggestion has already been put forward by Tarasov \citep[see][]{henrichs:2003}, which was at that time, ironically, rejected by one of the current authors. If this close companion were to turn out to be a Be
star, this would clearly mean a major step forwards in understanding the $\beta$ Cep system, and also remove the unfulfillable constraint on Be star models it poses now.

\subsection{The binary components}
The star $\beta$ Cep (V=3.2) has a visual companion (V=7.9) at a distance of 13.4\arcsec. A second companion was detected using speckle interferometry by \citet{gezari:1972} at a distance of $\sim$0.25\arcsec, which was later found to have a visual magnitude of V=6.6. The parameters of the close binary orbit have been determined from the variations in the pulsation period due to the light time effect and speckle interferometry by \citet[][see also \citealt{hadrava:1996}]{pigulski:1992}. When recent, additional speckle measurements \citep{hartkopf:2001} 
 are taken into account, the current position of the companion is at a distance of about 0.1\arcsec\ from the primary, at a position angle of 42\degr\ (in the NE) on the sky. From the mass ratio determined from the binary orbit, the companion has an estimated spectral type around B6-8.

As the target is very bright and the approximate orbit is known, the technique of spectro-astrometry is particularly well-suited to resolving the question of the origin of the H$\alpha$ emission. Spectro-astrometry measures the relative spatial position of spectral 
features from a long-slit spectrum \citep[see][and references therein]{bailey:1998,porter:2004}. If one star in an otherwise unresolved binary has, for
example, H$\alpha$ emission, the photocentre across the line perpendicular to the dispersion direction will shift towards that star.
So far the technique has mainly been used to detect close binary companions \citep[e.g.][]{bailey:1998,baines:2006}, but also the individual spectra of binaries with a separation down to tens of milliarcseconds (mas) can be obtained.

\section{Observations and data reduction}
\label{obs&data}
Longslit spectra of $\beta$ Cep were obtained with the ALFOSC spectrograph at the Nordic Optical Telescope (NOT) on La Palma.
We used grism \#17 (2400 l/mm VPH), which gives a dispersion of 0.25 \AA/pixel for the $\sim$6350--6850 \AA\ range. The 1.9\arcsec\ off-centre slit was used to avoid a ghost near H$\alpha$. We observed with a typical seeing of $\sim$1.1\arcsec, resulting in an effective resolution of R$\approx$4500. The CCD with 2048x2048 pixels gives a spatial resolution of 0.19\arcsec/pixel, thereby giving a good sampling of the spatial profile of the spectrum.

A total of 11 spectra were obtained on 28 August 2006, between 5:40 and 5:53 UT (HJD 2453975.74), with exposure times between 2 and 5 sec. The star was positioned at three different locations on the slit, to check for possible instrumental effects. The angle of the slit on the sky was set to 42\degr\ (NE), which was confirmed by images obtained without the slit, leaving the orientation of the sky unchanged with this instrument.

Data were reduced using the IRAF software package. The CCD-frames were corrected for the bias level and divided by a normalised flatfield. Scattered light was subtracted. Wavelength calibration spectra were obtained using an Ne lamp. The resulting two-dimensional spectra were fitted by a Gaussian profile in the spatial direction at each wavelength step with the fitprofs routine, using a 5-point running average in the dispersion direction (comparable to the spectral resolution). We have checked that similar results were obtained when no correction for scattered light was applied, or when Voigt instead of Gaussian profiles were used. Further consistency checks were carried out by comparing the results for all individual spectra taken at different slit positions. All traces were similar to each other, strongly suggesting that instrumental artefacts are not present.

\begin{figure}[t!bph]
\begin{center}
\includegraphics[height=\linewidth, angle=-90]{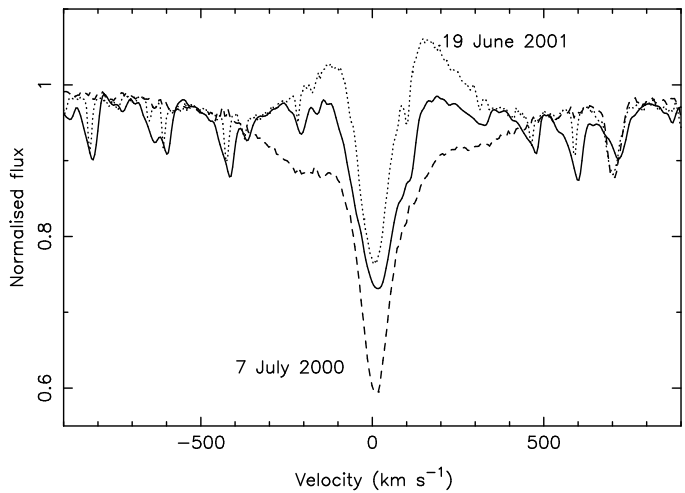}
\caption[]{Average H$\alpha$ profile (full line), the profile of 7 July 2000 and 19 June 2001 \citep[dashed and dotted lines respectively, see][]{henrichs:2003}. During our observations more emission was present than in 2000.}
\label{oldspec}
\end{center}
\end{figure}

\section{Results}
The average H$\alpha$ profile is plotted in Fig.~\ref{oldspec}, together with spectra taken in 2000 and 2001 \citep{henrichs:2003}. Although it is not directly clear from the new spectra that emission is present, comparison with the spectrum of July 2000 shows that the emission is currently stronger than it was in 2000.

\subsection{The source of the H$\alpha$ emission}
\begin{figure*}[btph]
\caption[]{Spectro-astrometric observations of H$\alpha$ (left) and the \ion{He}{i} line at 6678 \AA\ (right) of the $\beta$ Cep system. Shown are the normalised intensity line profile (top) and the position of the photocentre of the spatial profile relative to that of the continuum (bottom). In the plot of the offset of the photocentre the results of all individual spectra are shown (dotted lines) as well as the average (full line). In both the H$\alpha$ and \ion{He}{i} plots a shift of the photocentre towards the companion is visible. However, in H$\alpha$ the photocentre is offset from about $-$400 to +400 km s$^{-1}$, while in \ion{He}{i} the width of the offset is similar to the width of the line.}
\label{spectra}
\centering
\includegraphics[height=0.9\linewidth, angle=-90]{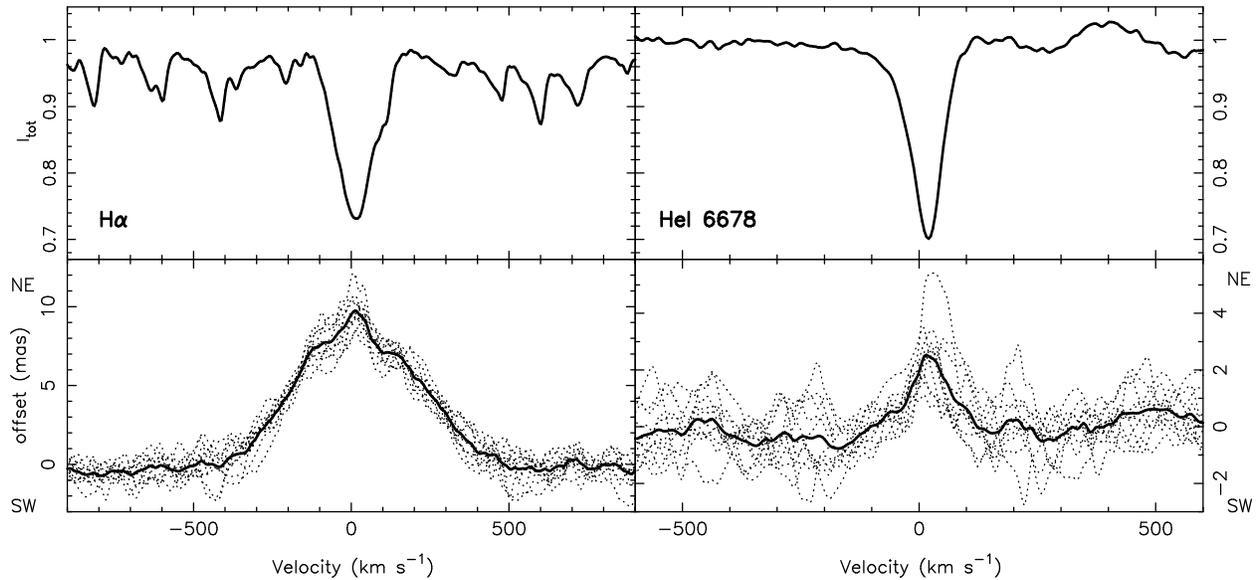}
\end{figure*}

The spectro-astrometric results for H$\alpha$ (6563 \AA) and the \ion{He}{i} line at 6678 \AA\ are shown in Fig.~\ref{spectra}. It is clear that near H$\alpha$ the photocentre of the spatial profile of the spectrum shifts towards the companion (in the NE direction). This is due to an increased relative contribution to the flux of the companion, indicating that the companion is the source of the H$\alpha$ emission. The width of the signature in H$\alpha$ is from about $-$400 to +400 km s$^{-1}$, which is much broader than the width of the absorption line and is typical for a Be star emission line.

\subsection{The spectra of the individual stars}
For close binaries with smaller separations than the slit, it is possible to determine the two spectra of the individual, unresolved, binary components with the technique described in \citet{bailey:1998SPIE} and \citet[][see also \citealt{porter:2004}]{takami:2003}. Using average photocentre shifts, we determined the individual spectra, adopting the measured magnitude difference of 3.4 magnitudes \citep{hartkopf:2001} and separations of 0.07\arcsec, 0.1\arcsec, and 0.15\arcsec, bracketing the estimated separation. 

\begin{figure*}[t!bph]
\caption[]{The results of the separation of the spectra of the primary and secondary components. We show the normalised intensity line profile (top) and the separated line profiles of the primary (middle) and the secondary (bottom). For the splitting of the spectra we have assumed a separation of 0.07\arcsec\ (dashed lines), 0.1\arcsec\ (full line, corresponding to the best estimate of the separation), and 0.15\arcsec\ (dotted line). A double-peaked H$\alpha$ emission line with a width of $\sim$400 km s$^{-1}$, typical of a classical Be star, is found in the secondary star. The \ion{He}{i} line at 6678 \AA\ is only present in the primary star.}
\label{spectrasplit}
\centering
\includegraphics[height=0.9\linewidth, angle=-90]{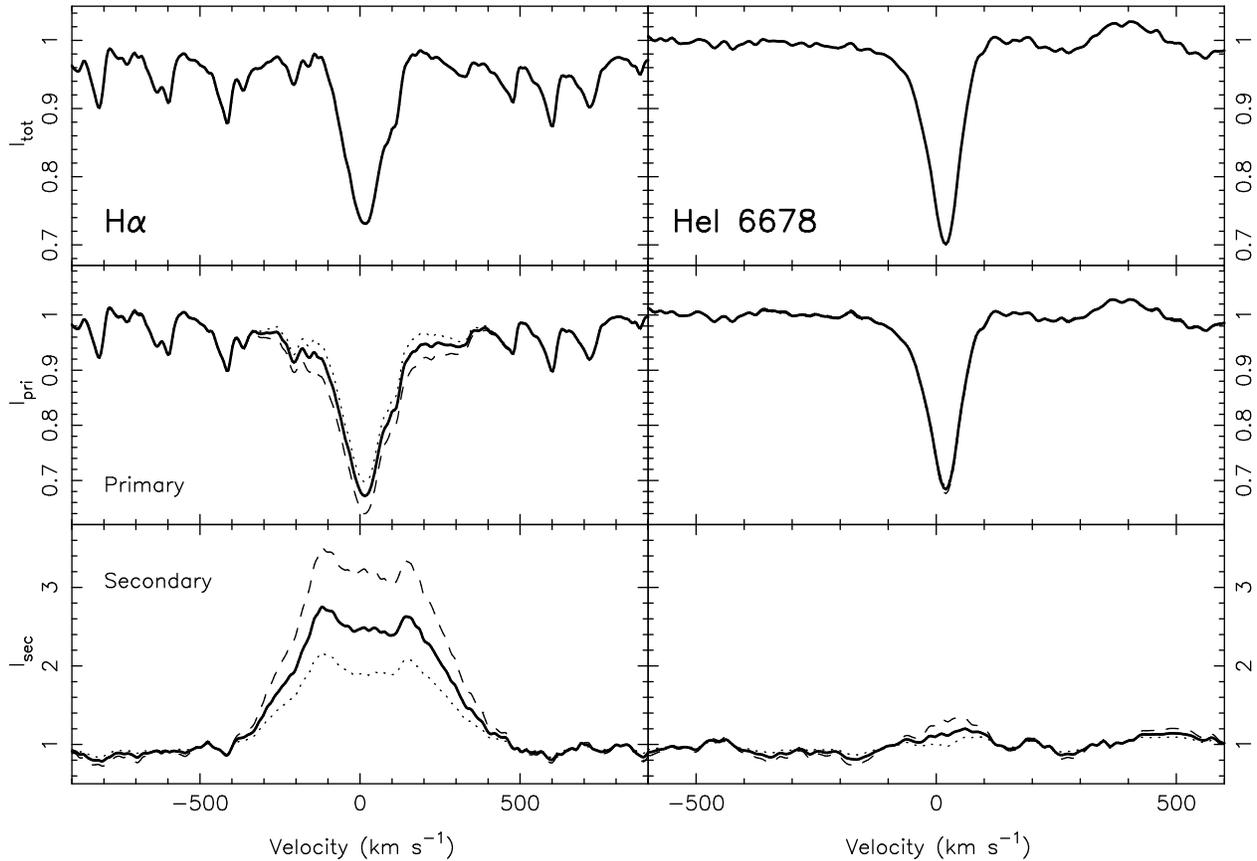}
\end{figure*}

The resulting spectra are shown in Fig.~\ref{spectrasplit}. The results for three possible separations are shown, and apart from the strength of H$\alpha$ the results are qualitatively similar. The conclusion that the NE component is the source of the H$\alpha$ emission is confirmed when the spectra are split. We find that the secondary has a double-peaked emission profile, characteristic of a classical Be star.

In the \ion{He}{i} line the signal is also in the direction of the companion, but it has the same width as the absorption line in the total intensity spectrum. In the separated spectra it can be seen that this line is present only in the primary and not in the secondary, as expected for its later spectral type.

\section{Conclusions and discussion}
\label{conclusions}
We have shown that the H$\alpha$ emission observed from the $\beta$ Cep system is not related to the slowly rotating primary star, but to the secondary, which is most likely a classical Be star. This explains why the H$\alpha$ emission is not modulated by the rotation of the primary. This removes the exceptional status of $\beta$ Cep among the fast rotating Be stars, which therefore no longer contradicts the current models that require rapid rotation for explaining the Be phenomenon.

We find that the H$\alpha$ emission extends from about $-$400 to +400 km~s$^{-1}$, in agreement with the results from \citet{hadrava:1996} and \citet{panko:1997}. This is independently confirmed by the extent of the variability shown in Fig.~\ref{oldspec}. The large width of the H$\alpha$ emission suggests a relatively high value for $v \sin i$, which points to a high inclination angle. With the orbital inclination angle of 87\degr\ \citep{pigulski:1992} and the high inclination angle of $\beta$ Cep itself \citep[$>$60\degr,][]{telting:1997,donati:2001} this means that the spin and orbital angular momentum vectors could well be aligned. An interesting question is how such a binary system with one, presumably spun-down, magnetic B star and a Be star may have evolved.

Our result implies that the observed H$\alpha$ emission is not related to the magnetic field of the primary star. This agrees with models explaining the variability observed in the UV wind-lines as due to the rotation of the magnetic field.

New spectro-astrometric observations to obtain a wider spectral coverage are being planned and will allow us to further constrain the $v \sin i$ and spectral type of the secondary star.

{\acknowledgements Based on observations obtained with the Nordic Optical Telescope, which is operated on the island of La Palma jointly by Denmark, Finland, Iceland, Norway, and Sweden, at the Spanish Observatorio del Roque de los Muchachos of the Instituto de Astrofisica de Canarias. The data presented here were taken using ALFOSC, which is owned by the Instituto de Astrofisica de Andalucia (IAA) and operated at the Nordic Optical Telescope under agreement between IAA and the NBIfAFG of the Astronomical Observatory of Copenhagen. RS and HFH thank F.\ Leone for useful discussions.}

\bibliographystyle{aa}
\bibliography{../../references}
\end{document}